\documentclass[aps,prl,twocolumn,amsmath,amssymb,nofootinbib,superscriptaddress]{revtex4-1}
\usepackage{times}
\usepackage[pdftex]{graphicx}
\usepackage{dcolumn}
\usepackage{bm}
\usepackage{amsmath}
\usepackage{indentfirst}
\usepackage{float}
\usepackage[colorlinks]{hyperref}
\usepackage[dvipsnames]{xcolor}
\usepackage{xcolor}
\newcommand{\hhl}[1]{{\color{black}{#1}}}

\colorlet{shadecolor}{gray!40}

\usepackage{subfigure}
\usepackage{algorithm}
\usepackage{algorithmicx}
\usepackage{algpseudocode}
\usepackage{amsmath}
\usepackage{verbatim}
\usepackage{makecell}
\usepackage[normalem]{ulem}
\usepackage{setspace}

\begin{document}

\title{Logical Magic State Preparation with Fidelity Beyond the Distillation Threshold on a Superconducting Quantum Processor}

\author{Yangsen Ye}
\thanks{These three authors contributed equally}
\author{Tan He}
\thanks{These three authors contributed equally}
\affiliation{Hefei National Research Center for Physical Sciences at the Microscale and School of Physical Sciences, University of Science and Technology of China, Hefei 230026, China}
\affiliation{Shanghai Research Center for Quantum Science and CAS Center for Excellence in Quantum Information and Quantum Physics, University of Science and Technology of China, Shanghai 201315, China}
\author{He-Liang Huang}
\thanks{These three authors contributed equally}
\affiliation{Hefei National Research Center for Physical Sciences at the Microscale and School of Physical Sciences, University of Science and Technology of China, Hefei 230026, China}
\affiliation{Shanghai Research Center for Quantum Science and CAS Center for Excellence in Quantum Information and Quantum Physics, University of Science and Technology of China, Shanghai 201315, China}
\affiliation{Henan Key Laboratory of Quantum Information and Cryptography, Zhengzhou, Henan 450000, China}
\author{Zuolin Wei}
\author{Yiming Zhang}
\author{Youwei Zhao}
\author{Dachao Wu}
\affiliation{Hefei National Research Center for Physical Sciences at the 
Microscale and School of Physical Sciences, University of Science and 
Technology of China, Hefei 230026, China}
\affiliation{Shanghai Research Center for Quantum Science and CAS Center for 
Excellence in Quantum Information and Quantum Physics, University of Science 
and Technology of China, Shanghai 201315, China}

\author{Qingling Zhu}
\affiliation{Shanghai Research Center for Quantum Science and CAS Center for 
Excellence in Quantum Information and Quantum Physics, University of Science 
and Technology of China, Shanghai 201315, China}
\affiliation{Hefei National Laboratory, University of Science and Technology of 
China, Hefei 230088, China}

\author{Huijie Guan}
\author{Sirui Cao}
\affiliation{Hefei National Research Center for Physical Sciences at the 
Microscale and School of Physical Sciences, University of Science and 
Technology of China, Hefei 230026, China}
\affiliation{Shanghai Research Center for Quantum Science and CAS Center for 
Excellence in Quantum Information and Quantum Physics, University of Science 
and Technology of China, Shanghai 201315, China}

\author{Fusheng Chen}
\author{Tung-Hsun Chung}
\affiliation{Shanghai Research Center for Quantum Science and CAS Center for 
Excellence in Quantum Information and Quantum Physics, University of Science 
and Technology of China, Shanghai 201315, China}
\affiliation{Hefei National Laboratory, University of Science and Technology of 
China, Hefei 230088, China}

\author{Hui Deng}
\affiliation{Hefei National Research Center for Physical Sciences at the 
Microscale and School of Physical Sciences, University of Science and 
Technology of China, Hefei 230026, China}
\affiliation{Shanghai Research Center for Quantum Science and CAS Center for 
Excellence in Quantum Information and Quantum Physics, University of Science 
and Technology of China, Shanghai 201315, China}
\affiliation{Hefei National Laboratory, University of Science and Technology of 
China, Hefei 230088, China}

\author{Daojin Fan}
\affiliation{Hefei National Research Center for Physical Sciences at the 
Microscale and School of Physical Sciences, University of Science and 
Technology of China, Hefei 230026, China}
\affiliation{Shanghai Research Center for Quantum Science and CAS Center for 
Excellence in Quantum Information and Quantum Physics, University of Science 
and Technology of China, Shanghai 201315, China}

\author{Ming Gong}
\affiliation{Hefei National Research Center for Physical Sciences at the 
Microscale and School of Physical Sciences, University of Science and 
Technology of China, Hefei 230026, China}
\affiliation{Shanghai Research Center for Quantum Science and CAS Center for 
Excellence in Quantum Information and Quantum Physics, University of Science 
and Technology of China, Shanghai 201315, China}
\affiliation{Hefei National Laboratory, University of Science and Technology of 
China, Hefei 230088, China}

\author{Cheng Guo}
\author{Shaojun Guo}
\author{Lianchen Han}
\author{Na Li}
\affiliation{Hefei National Research Center for Physical Sciences at the 
Microscale and School of Physical Sciences, University of Science and 
Technology of China, Hefei 230026, China}
\affiliation{Shanghai Research Center for Quantum Science and CAS Center for 
Excellence in Quantum Information and Quantum Physics, University of Science 
and Technology of China, Shanghai 201315, China}

\author{Shaowei Li}
\affiliation{Shanghai Research Center for Quantum Science and CAS Center for 
Excellence in Quantum Information and Quantum Physics, University of Science 
and Technology of China, Shanghai 201315, China}
\affiliation{Hefei National Laboratory, University of Science and Technology of 
China, Hefei 230088, China}

\author{Yuan Li}
\affiliation{Hefei National Research Center for Physical Sciences at the 
Microscale and School of Physical Sciences, University of Science and 
Technology of China, Hefei 230026, China}
\affiliation{Shanghai Research Center for Quantum Science and CAS Center for 
Excellence in Quantum Information and Quantum Physics, University of Science 
and Technology of China, Shanghai 201315, China}

\author{Futian Liang}
\affiliation{Hefei National Research Center for Physical Sciences at the 
Microscale and School of Physical Sciences, University of Science and 
Technology of China, Hefei 230026, China}
\affiliation{Shanghai Research Center for Quantum Science and CAS Center for 
Excellence in Quantum Information and Quantum Physics, University of Science 
and Technology of China, Shanghai 201315, China}
\affiliation{Hefei National Laboratory, University of Science and Technology of 
China, Hefei 230088, China}

\author{Jin Lin}
\affiliation{Shanghai Research Center for Quantum Science and CAS Center for 
Excellence in Quantum Information and Quantum Physics, University of Science 
and Technology of China, Shanghai 201315, China}
\affiliation{Hefei National Laboratory, University of Science and Technology of 
China, Hefei 230088, China}

\author{Haoran Qian}
\author{Hao Rong}
\author{Hong Su}
\author{Shiyu Wang}
\author{Yulin Wu}
\affiliation{Hefei National Research Center for Physical Sciences at the 
Microscale and School of Physical Sciences, University of Science and 
Technology of China, Hefei 230026, China}
\affiliation{Shanghai Research Center for Quantum Science and CAS Center for 
Excellence in Quantum Information and Quantum Physics, University of Science 
and Technology of China, Shanghai 201315, China}

\author{Yu Xu}
\affiliation{Shanghai Research Center for Quantum Science and CAS Center for 
Excellence in Quantum Information and Quantum Physics, University of Science 
and Technology of China, Shanghai 201315, China}
\affiliation{Hefei National Laboratory, University of Science and Technology of 
China, Hefei 230088, China}

\author{Chong Ying}
\author{Jiale Yu}
\affiliation{Hefei National Research Center for Physical Sciences at the 
Microscale and School of Physical Sciences, University of Science and 
Technology of China, Hefei 230026, China}
\affiliation{Shanghai Research Center for Quantum Science and CAS Center for 
Excellence in Quantum Information and Quantum Physics, University of Science 
and Technology of China, Shanghai 201315, China}

\author{Chen Zha}
\affiliation{Shanghai Research Center for Quantum Science and CAS Center for 
Excellence in Quantum Information and Quantum Physics, University of Science 
and Technology of China, Shanghai 201315, China}
\affiliation{Hefei National Laboratory, University of Science and Technology of 
China, Hefei 230088, China}

\author{Kaili Zhang}
\affiliation{Shanghai Research Center for Quantum Science and CAS Center for 
Excellence in Quantum Information and Quantum Physics, University of Science 
and Technology of China, Shanghai 201315, China}

\author{Yong-Heng Huo}
\author{Chao-Yang Lu}
\author{Cheng-Zhi Peng}
\author{Xiaobo Zhu}
\author{Jian-Wei Pan}
\affiliation{Hefei National Research Center for Physical Sciences at the Microscale and School of Physical Sciences, University of Science and Technology of China, Hefei 230026, China}
\affiliation{Shanghai Research Center for Quantum Science and CAS Center for Excellence in Quantum Information and Quantum Physics, University of Science and Technology of China, Shanghai 201315, China}
\affiliation{Hefei National Laboratory, University of Science and Technology of China, Hefei 230088, China}



\begin{abstract}
Fault-tolerant quantum computing based on surface code has emerged as an attractive candidate for practical large-scale quantum computers to achieve robust noise resistance. To achieve universality, magic states preparation is a commonly approach for introducing non-Clifford gates. Here, we present a hardware-efficient and scalable protocol for arbitrary logical state preparation for the rotated surface code, and further experimentally implement it on the \textit{Zuchongzhi} 2.1 superconducting quantum processor. An average of \hhl{$0.8983 \pm 0.0002$} logical fidelity at different logical states with distance-three is achieved, \hhl{taking into account both state preparation and measurement errors.} In particular, \hhl{the magic states $|A^{\pi/4}\rangle_L$, $|H\rangle_L$, and $|T\rangle_L$ are prepared non-destructively with logical fidelities of $0.8771 \pm 0.0009 $, $0.9090 \pm 0.0009 $, and $0.8890 \pm 0.0010$, respectively, which are higher than the state distillation protocol threshold, 0.859 (for H-type magic state) and 0.827 (for T -type magic state).} Our work provides a viable and efficient avenue for generating  high-fidelity raw logical magic states, which is essential for realizing non-Clifford logical gates in the surface code. 

\end{abstract}

\maketitle

\section{Introduction}
Practical quantum computers are extremely difficult to engineer and build, as they are easily crippled by the inevitable noise in realistic quantum hardwares~\cite{huang2020superconducting,huang2023near}. Fault-tolerant quantum computing build on quantum error correction (QEC) offers a promising path to quantum computation at scale, by encoding the quantum information into logical qubits. In the past decades, much progress has been made to construct QEC schemes and realize QEC in the specific context of trapped ions~\cite{schindler2011experimental,egan2021fault,ryan2021realization,hilder2022fault}, superconducting circuits~\cite{andersen2020repeated,ai2021exponential,marques2021logical,zhao2022, krinner2022realizing,acharya2022suppressing,reed2012realization,ofek2016extending,corcoles2015demonstration,huang2021emulating}, photons~\cite{yao2012experimental,pittman2005demonstration, luo2021quantum, liu2019demonstration} and nitrogen-vacancy centers~\cite{waldherr2014quantum,taminiau2014universal,cramer2016repeated}. The surface code~\cite{kitaev2003fault,raussendorf2007fault,fowler2012surface}, a planar realization of Kitaev's toric code, is experimentally attractive as it requires only a two-dimensional lattice of qubits with nearestneighbour coupling, and has a high error threshold of about 1\%. The quality properties and great potential of surface code have driven efforts to scale up experiments from distance-two~\cite{andersen2020repeated,ai2021exponential,marques2021logical} to distance-three~\cite{zhao2022, krinner2022realizing} and even distance-five~\cite{acharya2022suppressing}, until reaching a practical level.

Working with logical qubits to achieve a specific computational task introduces additional overhead for logical quantum gate operations. The surface code provides a relatively low-overhead implementation of the logical Clifford gate. However, a quantum circuit consisting only Clifford gates is not computationally universal, nor does it confer no quantum computational advantage, since it can be efficiently simulated by classical computing~\cite{gottesman1998heisenberg,bu2019efficient}. In order to achieve computational universality, there must be at least one non-Clifford gate, such as $T$ gate. These non-Clifford gate can be implemented through magic state injection~\cite{bravyi2005universal,reichardt2005quantum,bravyi2012magic,o2017quantum,hastings2018distillation}, but unfortunately, it takes a large overhead and a huge number of magic state~\cite{fowler2018low,bravyi2012magic,campbell2017unified,o2017quantum}. Thus, fast and high-fidelity logical magic state preparation~\cite{li2015magic,lodyga2015simple,gidney2023cleaner} is crucial in the implementation of universal logical quantum gates.

\begin{figure*}[!htbp]
\begin{center}
\includegraphics[width=1\linewidth]{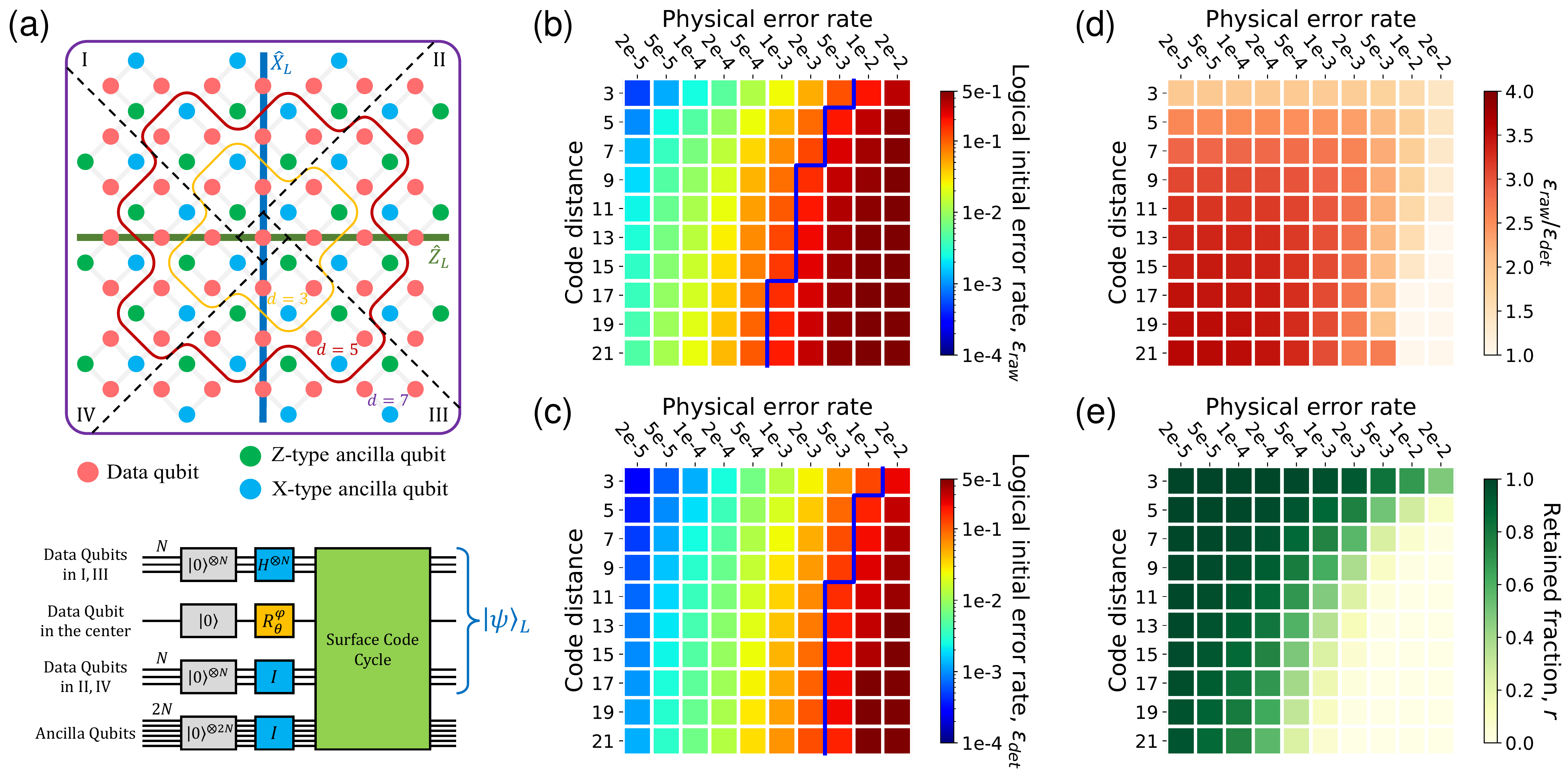}
\end{center}
\setlength{\abovecaptionskip}{0pt}
\caption{\textbf{Arbitrary logical state preparation protocol and simulation results.} \textbf{(a)} Arbitrary logical state preparation protocol. \textbf{Top panel:} The surface code is divided into 5 regions, the central data qubit, regions I, II, III, and IV. The logical operators $\hat{Z}_L$ and $\hat{X}_L$ intersect at the central data qubits. \textbf{Bottom panel:} The circuit of the protocol. All qubits are reset to $|0\rangle$ state at the beginning of the circuit. Then the  data qubits in the regions I and III are prepared to $|+\rangle$ by Hadamard gate, and the central data qubit is prepared to the target state $|\psi\rangle$ by rotation gates. One round of surface code cycle is applied afterwards, projecting the data qubits state into the logical state space. \textbf{(b-e)} Simulation results for the $|{+i}\rangle_L$ state preparation. \textbf{(b-c)} Logical initial error rate as a function of average physical error rate and surface code distance with no post-processing (b) and with post-selection of only syndrome measurements (c). The blue lines in the figure indicate the most demanding threshold, 0.141 (the 15-to-1 state distillation protocol for $H$-type magic state), for state distillation protocol. \textbf{(d-e)} The ratio of $\varepsilon_{raw}$ to $\varepsilon_{det}$ and the retained fraction of post-selection as a function of average physical error rate and surface code distance.}
\label{fig1}
\end{figure*}

This work aims to explore how to prepare arbitrary logical state, especially magic states, quickly and with high fidelity. Specifically, an arbitrary logical state preparation protocol is proposed for the rotated surface code, inspired by some relevant works~\cite{li2015magic,lodyga2015simple}. The protocol does not require extra ancilla qubits and is almost identical to the standard surface code protocol, except that the quantum state needs to be prepared to a specific product state according to the target logical state during the initialization stage. Theoretical analysis show good scaling behavior of the protocol for high-fidelity large-scale logical quantum state preparation. Furthermore, we experimentally realize the protocol on the \textit{Zuchongzhi} 2.1 superconducting quantum system~\cite{wu2021strong,zhu2021quantum} to demonstrate its practical performance on real quantum devices. An average logical fidelity of \hhl{$0.8983 \pm 0.0002$} is achieved with post-selection using syndrome measurements for different prepared logical states, even in the presence of significant readout errors during measurement. Among them, two $H$-type logical magic states ${|A^{\pi/4}\rangle}_L$, ${|H\rangle}_L$ and one $T$-type logical magic state ${|T\rangle}_L$ are obtained with logical fidelities of \hhl{$0.8771 \pm 0.0009 $, $0.9090 \pm 0.0009 $, and $0.8890 \pm 0.0010 $}, respectively. These are significantly higher than the 15-to-1 magic state distillation protocol threshold 0.859 (for $H$-type magic state) and the 5-to-1 magic state distillation protocol threshold 0.827 (for $T$-type magic state)~\cite{bravyi2005universal}. The achieved results suggest that our work represents a key step towards universal and scalable fault-tolerant quantum computing, and has the potential to play a crucial role in some NISQ protocols/algorithms~\cite{rolander2022quantum}, such as error mitigation~\cite{hicks2022active}.

\section{Arbitrary Logical State Preparation Protocol}
\vspace{-0.3cm}
The arbitrary logical state preparation protocol is shown in Fig.~\ref{fig1}(a). The basic idea is to initialize the data qubits to a specific quantum state first, and then apply one round surface code cycle to project the data qubits into the logical state space. Assume the target logical state is $|\psi\rangle_L=\alpha|0\rangle_L+\beta|1\rangle_L$, the detailed steps of our protocol can be described in the following:

\hspace{1ex}1. Reset all qubits to $|0\rangle$ state, including data qubits and ancilla qubits.

\hspace{1ex}2. Divide the rotated surface code into 5 regions, the central data qubit, regions I, II, III, and IV, as shown in the top panel of Fig.~\ref{fig1}(a). Prepare the data qubits in regions I and III to the $|+\rangle$ state and the data qubit in the center (intersection of logical operators $\hat{Z}_L$ and $\hat{X}_L$) to the target state $|\psi\rangle=\alpha|0\rangle+\beta|1\rangle$, while the data qubits of the remaining regions II and IV stay in the $|0\rangle$ state. The data qubits state after Step 2 is
\begin{align}
|\Psi_0\rangle = |\psi\rangle \bigotimes_{D_i\in \text{I} \cup \text{III}}|+\rangle  \bigotimes_{D_j\in \text{II} \cup \text{IV}}|0\rangle 
\end{align}
where $D_i\in \text{I} \cup \text{III}$ ($D_j\in \text{II} \cup \text{IV}$) is denoted as the data qubits of region I and III (II and IV), $D_c$ is denoted as the central data qubit.

\hspace{1ex}3. Apply one round of surface code cycle. After measuring all the ancilla qubits, the data qubits are then prepared to the desired logical state $|\psi\rangle_L$.

The protocol is applicable to the surface code with arbitrary distance $d$, and its quantum circuit is shown in the bottom panel of Fig.~\ref{fig1}(a). The $X$-stabilizers in regions I and III and the $Z$-stabilizers in regions II and IV would have the deterministic measurement values $0$ if no error occurred. With post-selection procedures using these stabilizers, the logical initial error rate can be effectively reduced.

We further investigate the performance of the protocol through numerical simulation (see Fig.~\ref{fig1}(b-e)). The simulation uses the Pauli depolarizing model, and uses the average physical error rate, \textit{i.e.} the error rate of all operations, including single-qubit gate, two-qubit gate, readout, reset and thermal excitation, are all the same. The target logical state in our simulations is chosen as $|{+i}\rangle_L$, and by doing so, the error detection capabilities of $X$- and $Z$-stabilizers can be tested simultaneously.

We simulate the logical initial error rate with different surface code distances and average physical error rates. The simulated circuits contain only one round of surface code cycle (as shown in Fig.~\ref{fig1}(a)), followed by a logical $Y$ measurement (see Supplemental Materials for details). The results without any post-processing and with syndrome measurement post-selection are shown in Fig.~\ref{fig1}(b) and (c), respectively. The blue line indicates the threshold for state distillation protocol of the magic state, the threshold here is chosen to be the most demanding 15-to-1 $H$-type protocol threshold of 0.141. As we can see, the logical initialization error rate increase with the increasing of the surface code distance and the average physical error rate. It is clear that there is a significant decrease in the logical error rate by the post-selection, and the blue line is thus moved towards the right, which means that more relaxed conditions are able to perform the state distillation procedure. Figure~\ref{fig1}(d) shows the ratio of logical initial error rate before and after post-selection, which represents the suppression ability of the post-selection for errors in the logical state preparation process. It can be seen that the suppression rate ranges from 1 to 4 times for different conditions, and the suppression ability becomes stronger as the code distance grows and the average physical error rate decreases.

The retained fraction of post-selection is shown in Fig.~\ref{fig1}(e). This is an important indicator for the the efficiency of state distillation. The retained fraction increases with decreasing code spacing and average physical error rate. When the average error rate is 0.0002, the retained fraction is acceptable in state distillation even for a code distance-21 (about $56\%$).

\begin{figure}[!tbp]
\begin{center}
\includegraphics[width=1.0\linewidth]{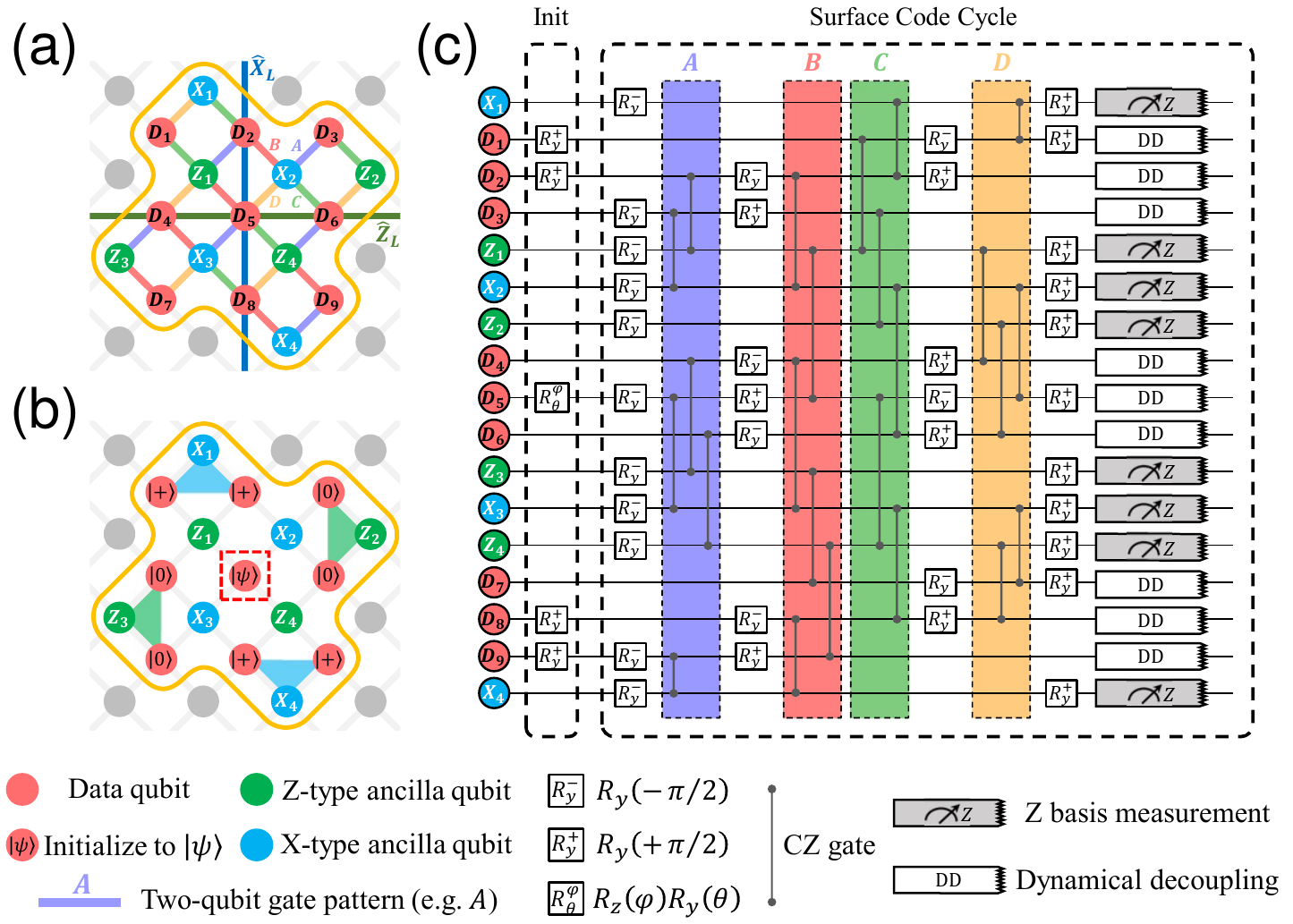}
\end{center}
\setlength{\abovecaptionskip}{0pt}
\caption{\textbf{Layout and circuit implementation.} \textbf{(a)} Structure of distance-three surface code, with 9 data qubits(red dots), 4 $Z$-type ancilla qubits(green dots) and 4 $X$-type ancilla qubits(blue dots). Connecting lines are colored according to their involvement in two-qubit gate layers as shown in (c). \textbf{(b)} Preparing 9 data qubits to a specific state, with 4 qubits stay in $|0\rangle$ state, 4 qubits initialize to $|+\rangle$ state and only one qubit transform to the target $|\psi\rangle$ state. \textbf{(c)} Circuit for preparing arbitrary logical state. First initialize 9 data qubits to specific states as (b) illustrated, then apply one round surface code cycle. Squares with different tags represent different single-qubit gates. All gates in one color block are applied simultaneously.}
\label{fig2}
\end{figure}

\section{Experimental Implementation on a superconducting quantum processor}

To demonstrate the performance of the protocol on a real quantum device, we create a distance-three surface code using 17 out of the 66 qubits on the \textit{Zuchongzhi} 2.1 superconducting quantum system (see Supplemental Materials for the system performance). This 17-qubit distance-three surface code (see Fig.~\ref{fig2}(a)) consists of 9 data qubits, 4 $X$-type ancilla qubits and 4 $Z$-type ancilla qubits. To prepare the logical state $|\psi\rangle_L=\cos{\left(\frac{\theta}{2}\right)} |0\rangle_L + e^{i\varphi} \sin{\left(\frac{\theta}{2}\right)} |1\rangle_L$, the data qubits are initialized in the way shown in Fig.~\ref{fig2}(b) to the product state
\begin{align} \label{eq:a1}
|\Psi\rangle = |+\rangle|+\rangle|0\rangle|0\rangle|\psi\rangle|0\rangle|0\rangle|+\rangle|+\rangle,
\end{align}
where $|\psi\rangle = \cos{\left(\frac{\theta}{2}\right)} |0\rangle + e^{i\varphi} \sin{\left(\frac{\theta}{2}\right)} |1\rangle$ can be experimentally realized using the virtual $Z$-gate and standard $\pi/2$ gate, as
\begin{align}
|\psi\rangle = Z_{\varphi} \cdot X_{\pi/2} \cdot Z_{\pi-\theta} \cdot X_{\pi/2}|0\rangle.
\end{align}
The corresponding quantum circuit is shown in Fig.~\ref{fig2}(c). After implementing one round of surface code cycle, the logical state $|\psi\rangle_L$ is prepared. Furthermore, as shown in Fig.~\ref{fig2}(b), in the logical state preparation process, 4 stabilizers $X_1$, $Z_2$, $Z_3$, $X_4$ are deterministic.

\begin{figure}[!htbp]
\begin{center}
\includegraphics[width=1.0\linewidth]{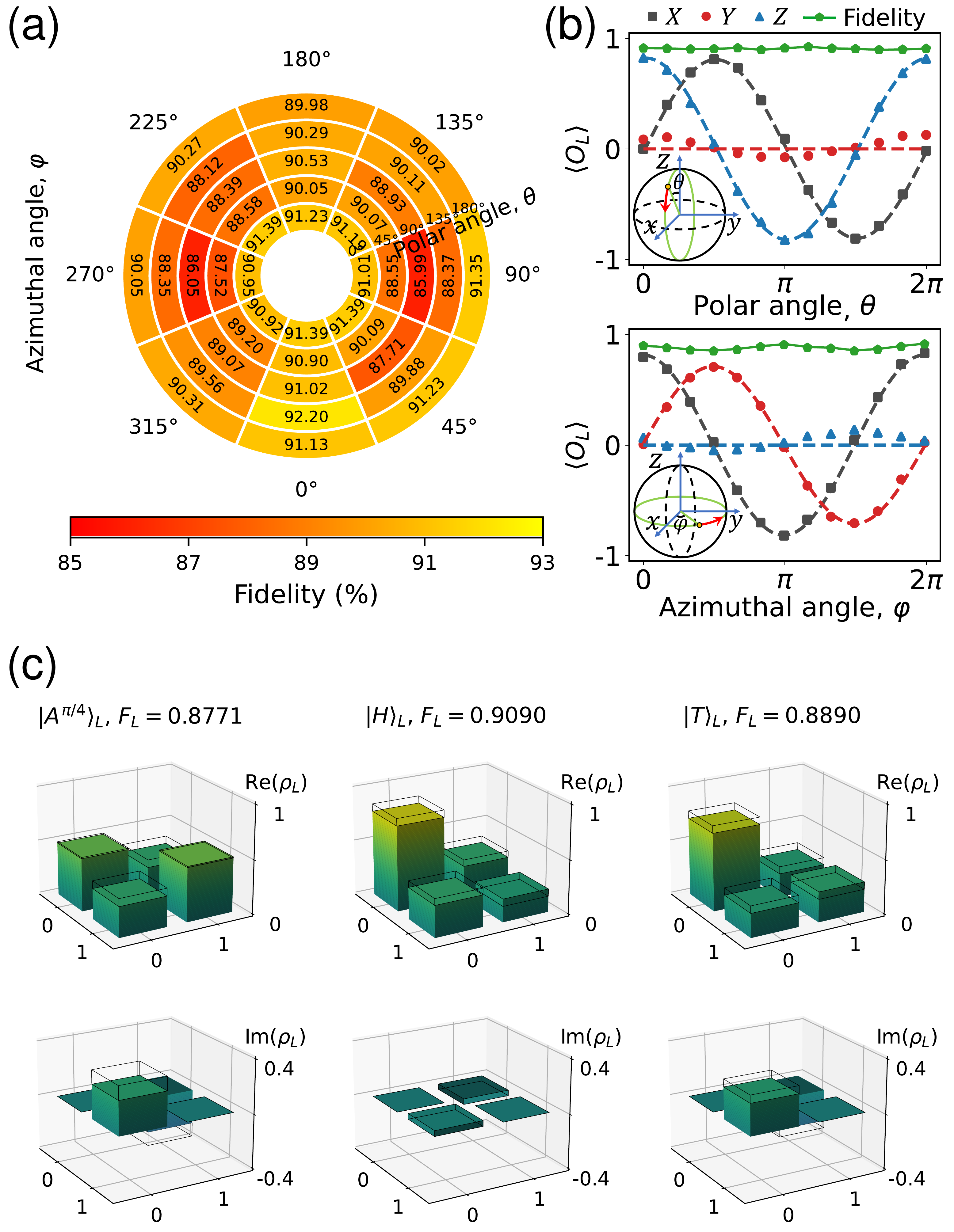}
\end{center}
\setlength{\abovecaptionskip}{0pt}
\caption{\textbf{Experimental results of the prepared different logical states.} \textbf{(a)} Logical state fidelity with post-selection in Bloch sphere. The fidelity of the preparation of different logical states is represented as a circle, which is divided into multiple annular sectors, each representing a point on the Bloch sphere, with the radial direction representing the polar angle $\theta$ and the tangential direction representing the azimuthal angle $\varphi$. The obtained average logical fidelity is \hhl{0.8983}. \textbf{(b)} Logical measurement results of $X_L$, $Y_L$, $Z_L$ as a function of polar angle $\theta$ or azimuthal angle $\varphi$. The colored dashed curves are the result of fitting with trigonometric function. \textbf{(c)} The logical density matrices of the magic states. Real and imaginary parts are represented separately, and the transparent wireframes represent the difference from the ideal density matrix.}
\label{fig3}
\end{figure}

We first prepare different logical states of uniformly scattered points on the Bloch sphere by varying the parameters of ${\theta}$ and ${\varphi}$. Figure~\ref{fig3}(a) shows the logical state fidelity $F_L$ of these prepared logical states after using post-selection to drop the results that have detection events during the preparation (see Supplemental Materials for the results before post-selection), where 
\begin{align}
F_L=\left(\text{Tr}\sqrt{\sqrt{\rho_{\text{exp}}}\cdot\rho_{\text{theory}}\cdot \sqrt{\rho_{\text{exp}}}}\right)^2,
\end{align}
and $\rho_{\text{exp}}$ is the experimental density matrix reconstructed by maximum-likelihood estimation after logical $X_L$, $Y_L$, $Z_L$ measurements. \hhl{We note that the measurement results include both state preparation and measurement (SPAM) error. We do not employ readout error mitigation strategies~\cite{smith2021qubit} to remove measurement errors because we believe it provides a more predictive assessment of the actual fidelity when generating and consuming magic states for a non-Clifford gate, as consuming the state involves measurement.} These fidelities are represented as pie-shaped (Fig.~\ref{fig3}(a)), which is divided into multiple annular sectors, each representing a point on the Bloch sphere, with the radial direction representing the polar angle $\theta$ and the tangential direction representing the azimuthal angle $\varphi$. The obtained average logical fidelity is $0.8983 \pm 0.0002$. Furthermore, we fixed one parameter in $\theta$ and $\varphi$, measuring the logical operators $\hat{X}_L$, $\hat{Y}_L$, $\hat{Z}_L$ to obtain expectation results as a function of the other parameter. As shown in Fig.~\ref{fig3}(b), the experimental points of $\langle X \rangle_L$, $\langle Y \rangle_L$ and $\langle Z \rangle_L$ are consistent with the sine/cosine variation.

Also, we show logical state tomography results of the prepared three magic states, including two $H$-type magic states ${|A^{\pi/4}\rangle}_L = \frac{1}{\sqrt{2}}\left({|0\rangle}_L + e^{i\pi/4}{|1\rangle}_L\right)$ and ${|H\rangle}_L = \cos{\frac{\pi}{8}}{|0\rangle}_L + \sin{\frac{\pi}{8}}{|1\rangle}_L$, and one $T$-type magic state ${|T\rangle}_L = \cos{\frac{\beta}{2}}{|0\rangle}_L + e^{i\pi/4}\sin{\frac{\beta}{2}}{|1\rangle}_L$, where $\beta=\arccos{\frac{1}{\sqrt{3}}}$. These two type magic states are the quantum resources for realizing non-Clifford gates $\Lambda({e^{ - i\pi /4}})$ and $\Lambda({e^{ - i\pi /6}})$, where
\begin{align}
\Lambda({e^{ - i\theta }}) = \left( {\begin{array}{*{20}{c}}
    1&0\\
    0&{{e^{i\theta }}}
    \end{array}} \right).
\end{align}

In Fig.~\ref{fig3}(c), the real and imaginary parts of the density matrix are shown separately. The logical fidelities of these magic states are \hhl{$|A^{\pi/4}\rangle_L$: $0.8771 \pm 0.0009 $, ${|H\rangle}_L$: $0.9090 \pm 0.0009 $, and ${|T\rangle}_L$: $0.8890 \pm 0.0010$}, which exceed the threshold of the respective state distillation.

\begin{figure*}[!htbp]
\begin{center}
\includegraphics[width=1.0\linewidth]{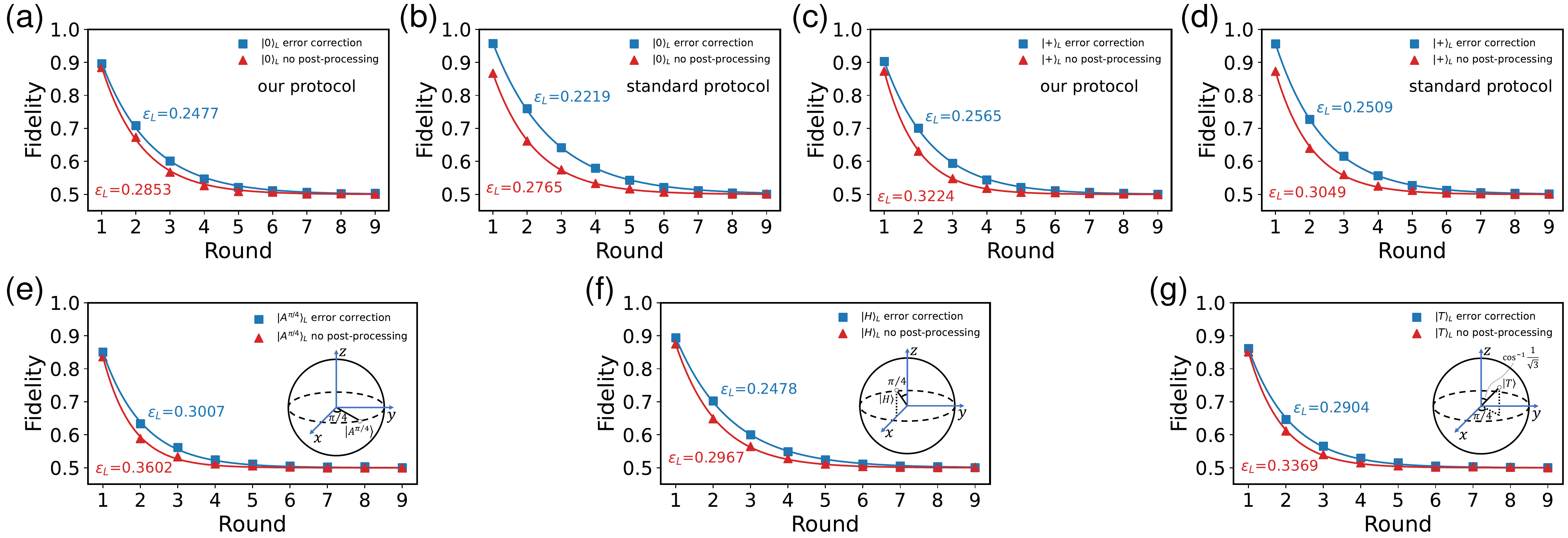}
\end{center}
\setlength{\abovecaptionskip}{0pt}
\caption{\textbf{Fidelity of different logical state with error correction.} \textbf{(a) and (b)} show the fidelity of logical $|0\rangle_L$ state with the number of surface code cycles with (blue line with square) and without (red line with triangular) error correction, by using our arbitrary logical state preparation proposal (a) and standard proposal (b), respectively. \textbf{(c)-(d)} Same as (a)-(b) with logical $|+\rangle_L$ state, with arbitrary logical state preparation proposal (c) and standard proposal (d). \textbf{(e), (f), and (g)} are results for $|A^{\pi/4}\rangle_L$, $|H\rangle_L$, and $|T\rangle_L$ state using our arbitrary logical state preparation proposal, respectively.}
\label{fig5}
\end{figure*}

To observe the error correction performance of the surface code for different logical initial states, we repeatedly apply the surface code cycles after the logical state is prepared. Fig.~\ref{fig5} shows how the fidelity the logical states varies with the number of surface code cycles with and without error correction. The logical error rates are derived by fitting the curves with $\mathcal{F}_L(k) = \frac{1}{2} \left(1+(1-2 \epsilon_L)^{k-k_0}\right)$~\cite{o2017density}. Fig.~\ref{fig5}(a) and (c) show the results for $|0\rangle_L$ and $|+\rangle_L$ using the arbitrary state preparation protocol. The logical error rates per round of $|0\rangle_L$ and $|+\rangle_L$ without error correction are $28.53\%$ and $32.24\%$. After the error correction procedure, the fidelity of the logical states at each point is improved and the logical error rates per round are reduced to $24.77\%$ and $25.65\%$, respectively. As a comparison, the results using the standard protocol~\cite{fowler2012surface} for $|0\rangle_L$ and $|+\rangle_L$ are shown in Fig.~\ref{fig5}(b) and (d), and we can observe that the logical error rates per round of the two protocols are similar. The most obvious difference is that in the results obtained by standard protocol, the fidelity of the logical states at the first round is significantly improved with error correction. This is mainly because that all the four stabilizers $Z_1$, $Z_2$, $Z_3$, $Z_4$ ($X_1$, $X_2$, $X_3$, $X_4$) are work for the $|0\rangle_L$ ($|+\rangle_L$) in the first round, while only half of stabilizers, $X_1$ and $X_4$ ($Z_2$ and $Z_3$), work for the $|0\rangle_L$ ($|+\rangle_L$) in our arbitrary logical state preparation protocol (as shown in Fig.~\ref{fig2}(b)). These stabilizers are in the edge positions and not near-neighbors, so it is difficult to correct error during logical state preparation. The most valuable aspect of the arbitrary logical state preparation protocol is that it can simply prepare arbitrary logical states, whereas the standard approach requires very complex operations. Fig.~\ref{fig5} (e), (f), and (g) show the results for magic states ${|A^{\pi/4}\rangle}_L$, ${|H\rangle}_L$ and ${|T\rangle}_L$. The achieved results show that the logical error rates per round for these prepared complex logical states are comparable to that of the standard logical states.

\section{CONCLUSION AND OUTLOOK}

\hhl{The crucial step for surface code based fault-tolerant computing, preparing distance-three logical magic state with fidelity beyond the distillation threshold, is achieved in this work.} Our work provides a highly simple, experimentally friendly, and scalable way to prepare high-fidelity raw magic states, which is critical for decreasing the overhead for distillation, and thus paving the way for practical fault-tolerant quantum computing. The protocol developed is partially fault-tolerant and naturally compatible with the error detection and repeated error correction, to enhance the logical state fidelity as well as to lift the logical coherence time. It might be improved to fully fault-tolerant by introducing a flag qubit mechanism~\cite{chamberland2020very,chamberland2019fault}. In addition, using some new approaches may further enhance the fidelity of magic state preparation and measurement~\cite{gidney2023cleaner,gidney2023inplace}. All of these will be left for our future work.

\begin{acknowledgments}
The authors are grateful for valuable discussions with Craig Gidney and Ying Li. The authors thank the USTC Center for Micro- and Nanoscale Research and Fabrication for supporting the sample fabrication. The authors also thank QuantumCTek Co., Ltd., for supporting the fabrication and the maintenance of room-temperature electronics.
\textbf{Funding:}
This research was supported by the Chinese Academy of Sciences, Anhui 
Initiative in Quantum Information Technologies, Shanghai Municipal Science and 
Technology Major Project (Grant No. 2019SHZDZX01), Innovation Program for 
Quantum Science and Technology (Grant No. 2021ZD0300200), Special funds from 
Jinan science and Technology Bureau and Jinan high tech Zone Management 
Committee, Technology Committee of Shanghai Municipality, National Science 
Foundation of China (Grants No. 11905217, No. 11774326), Natural Science Foundation of Shandong Province, China (grant number
ZR202209080019), and Natural Science Foundation of Shanghai (Grant No. 23ZR1469600), the Shanghai Sailing Program (Grant No. 23YF1452600). X. B. Zhu acknowledges support 
from the New Cornerstone Science Foundation through the XPLORER PRIZE. H.-L. H. acknowledges support from the Youth Talent 
Lifting Project (Grant No. 2020-JCJQ-QT-030), National Natural Science 
Foundation of China (Grants No. 11905294, 12274464), China Postdoctoral Science 
Foundation, and the Open Research Fund from State Key Laboratory of High 
Performance Computing of China (Grant No. 201901-01).
\end{acknowledgments}

\bibliographystyle{apsrev4-1}
\bibliography{references}

\end{document}


\title{Supplemental Material for \\ ``Logical Magic State Preparation with Fidelity Beyond the Distillation Threshold on a Superconducting Quantum Processor''}

\author{Yangsen Ye}
\thanks{These three authors contributed equally}
\author{Tan He}
\thanks{These three authors contributed equally}
\affiliation{Hefei National Research Center for Physical Sciences at the Microscale and School of Physical Sciences, University of Science and Technology of China, Hefei 230026, China}
\affiliation{Shanghai Research Center for Quantum Science and CAS Center for Excellence in Quantum Information and Quantum Physics, University of Science and Technology of China, Shanghai 201315, China}
\author{He-Liang Huang}
\thanks{These three authors contributed equally}
\affiliation{Hefei National Research Center for Physical Sciences at the Microscale and School of Physical Sciences, University of Science and Technology of China, Hefei 230026, China}
\affiliation{Shanghai Research Center for Quantum Science and CAS Center for Excellence in Quantum Information and Quantum Physics, University of Science and Technology of China, Shanghai 201315, China}
\affiliation{Henan Key Laboratory of Quantum Information and Cryptography, Zhengzhou, Henan 450000, China}
\author{Zuolin Wei}
\author{Yiming Zhang}
\author{Youwei Zhao}
\author{Dachao Wu}
\affiliation{Hefei National Research Center for Physical Sciences at the 
Microscale and School of Physical Sciences, University of Science and 
Technology of China, Hefei 230026, China}
\affiliation{Shanghai Research Center for Quantum Science and CAS Center for 
Excellence in Quantum Information and Quantum Physics, University of Science 
and Technology of China, Shanghai 201315, China}

\author{Qingling Zhu}
\affiliation{Shanghai Research Center for Quantum Science and CAS Center for 
Excellence in Quantum Information and Quantum Physics, University of Science 
and Technology of China, Shanghai 201315, China}
\affiliation{Hefei National Laboratory, University of Science and Technology of 
China, Hefei 230088, China}

\author{Huijie Guan}
\author{Sirui Cao}
\affiliation{Hefei National Research Center for Physical Sciences at the 
Microscale and School of Physical Sciences, University of Science and 
Technology of China, Hefei 230026, China}
\affiliation{Shanghai Research Center for Quantum Science and CAS Center for 
Excellence in Quantum Information and Quantum Physics, University of Science 
and Technology of China, Shanghai 201315, China}

\author{Fusheng Chen}
\author{Tung-Hsun Chung}
\affiliation{Shanghai Research Center for Quantum Science and CAS Center for 
Excellence in Quantum Information and Quantum Physics, University of Science 
and Technology of China, Shanghai 201315, China}
\affiliation{Hefei National Laboratory, University of Science and Technology of 
China, Hefei 230088, China}

\author{Hui Deng}
\affiliation{Hefei National Research Center for Physical Sciences at the 
Microscale and School of Physical Sciences, University of Science and 
Technology of China, Hefei 230026, China}
\affiliation{Shanghai Research Center for Quantum Science and CAS Center for 
Excellence in Quantum Information and Quantum Physics, University of Science 
and Technology of China, Shanghai 201315, China}
\affiliation{Hefei National Laboratory, University of Science and Technology of 
China, Hefei 230088, China}

\author{Daojin Fan}
\affiliation{Hefei National Research Center for Physical Sciences at the 
Microscale and School of Physical Sciences, University of Science and 
Technology of China, Hefei 230026, China}
\affiliation{Shanghai Research Center for Quantum Science and CAS Center for 
Excellence in Quantum Information and Quantum Physics, University of Science 
and Technology of China, Shanghai 201315, China}

\author{Ming Gong}
\affiliation{Hefei National Research Center for Physical Sciences at the 
Microscale and School of Physical Sciences, University of Science and 
Technology of China, Hefei 230026, China}
\affiliation{Shanghai Research Center for Quantum Science and CAS Center for 
Excellence in Quantum Information and Quantum Physics, University of Science 
and Technology of China, Shanghai 201315, China}
\affiliation{Hefei National Laboratory, University of Science and Technology of 
China, Hefei 230088, China}

\author{Cheng Guo}
\author{Shaojun Guo}
\author{Lianchen Han}
\author{Na Li}
\affiliation{Hefei National Research Center for Physical Sciences at the 
Microscale and School of Physical Sciences, University of Science and 
Technology of China, Hefei 230026, China}
\affiliation{Shanghai Research Center for Quantum Science and CAS Center for 
Excellence in Quantum Information and Quantum Physics, University of Science 
and Technology of China, Shanghai 201315, China}

\author{Shaowei Li}
\affiliation{Shanghai Research Center for Quantum Science and CAS Center for 
Excellence in Quantum Information and Quantum Physics, University of Science 
and Technology of China, Shanghai 201315, China}
\affiliation{Hefei National Laboratory, University of Science and Technology of 
China, Hefei 230088, China}

\author{Yuan Li}
\affiliation{Hefei National Research Center for Physical Sciences at the 
Microscale and School of Physical Sciences, University of Science and 
Technology of China, Hefei 230026, China}
\affiliation{Shanghai Research Center for Quantum Science and CAS Center for 
Excellence in Quantum Information and Quantum Physics, University of Science 
and Technology of China, Shanghai 201315, China}

\author{Futian Liang}
\affiliation{Hefei National Research Center for Physical Sciences at the 
Microscale and School of Physical Sciences, University of Science and 
Technology of China, Hefei 230026, China}
\affiliation{Shanghai Research Center for Quantum Science and CAS Center for 
Excellence in Quantum Information and Quantum Physics, University of Science 
and Technology of China, Shanghai 201315, China}
\affiliation{Hefei National Laboratory, University of Science and Technology of 
China, Hefei 230088, China}

\author{Jin Lin}
\affiliation{Shanghai Research Center for Quantum Science and CAS Center for 
Excellence in Quantum Information and Quantum Physics, University of Science 
and Technology of China, Shanghai 201315, China}
\affiliation{Hefei National Laboratory, University of Science and Technology of 
China, Hefei 230088, China}

\author{Haoran Qian}
\author{Hao Rong}
\author{Hong Su}
\author{Shiyu Wang}
\author{Yulin Wu}
\affiliation{Hefei National Research Center for Physical Sciences at the 
Microscale and School of Physical Sciences, University of Science and 
Technology of China, Hefei 230026, China}
\affiliation{Shanghai Research Center for Quantum Science and CAS Center for 
Excellence in Quantum Information and Quantum Physics, University of Science 
and Technology of China, Shanghai 201315, China}

\author{Yu Xu}
\affiliation{Shanghai Research Center for Quantum Science and CAS Center for 
Excellence in Quantum Information and Quantum Physics, University of Science 
and Technology of China, Shanghai 201315, China}
\affiliation{Hefei National Laboratory, University of Science and Technology of 
China, Hefei 230088, China}

\author{Chong Ying}
\author{Jiale Yu}
\affiliation{Hefei National Research Center for Physical Sciences at the 
Microscale and School of Physical Sciences, University of Science and 
Technology of China, Hefei 230026, China}
\affiliation{Shanghai Research Center for Quantum Science and CAS Center for 
Excellence in Quantum Information and Quantum Physics, University of Science 
and Technology of China, Shanghai 201315, China}

\author{Chen Zha}
\affiliation{Shanghai Research Center for Quantum Science and CAS Center for 
Excellence in Quantum Information and Quantum Physics, University of Science 
and Technology of China, Shanghai 201315, China}
\affiliation{Hefei National Laboratory, University of Science and Technology of 
China, Hefei 230088, China}

\author{Kaili Zhang}
\affiliation{Shanghai Research Center for Quantum Science and CAS Center for 
Excellence in Quantum Information and Quantum Physics, University of Science 
and Technology of China, Shanghai 201315, China}

\author{Yong-Heng Huo}
\author{Chao-Yang Lu}
\author{Cheng-Zhi Peng}
\author{Xiaobo Zhu}
\author{Jian-Wei Pan}
\affiliation{Hefei National Research Center for Physical Sciences at the Microscale and School of Physical Sciences, University of Science and Technology of China, Hefei 230026, China}
\affiliation{Shanghai Research Center for Quantum Science and CAS Center for Excellence in Quantum Information and Quantum Physics, University of Science and Technology of China, Shanghai 201315, China}
\affiliation{Hefei National Laboratory, University of Science and Technology of China, Hefei 230088, China}

\date{\today}

\maketitle

\setcounter{section}{0}
\renewcommand{\thefigure}{S\arabic{figure}}	
\renewcommand{\thetable}{S\arabic{table}}	
\renewcommand{\theequation}{S\arabic{equation}}	
\setcounter{figure}{0}
\setcounter{table}{0}
\setcounter{equation}{0}

\section{System Calibration}

\begin{table*}[!htbp]
	\centering
	\begin{tabular}{c| c c c c c c c c c c}
		\hline
		\hline
		Parameters& X1& Z1& X2& Z2& Z3& X3& Z4& X4& &Average\\
		\hline
		\hline
		Qubit maximum frequency, $\omega_{\textrm{q}}^{\textrm{max}}/2\pi$
		(GHz)& 5.079& 4.989& 5.119& 5.109& 5.188& 5.144& 5.032& 5.284& & 5.118
\\
		\hline
		Qubit idle frequency, $\omega_{\textrm{q}}/2\pi$ (GHz)& 4.916& 4.920& 5.084& 5.031& 5.101& 4.956& 4.903& 5.186& & 5.012
\\
		\hline
		Readout drive frequency, $\omega_{\textrm{r}}/2\pi$ (GHz)& 6.416& 6.413& 6.351& 6.288& 6.464& 6.409& 6.346& 6.345& & 6.379
\\
		\hline
		Qubit anharmonicity, $U/2\pi $ (MHz)& -258& -242& -238& -246& -243& -242& -244& -250& & -246
\\
		\hline
		Energy relaxation time $T_1$ at idle frequency ($\mu$s)& 26.8& 33.5& 23.6& 22.7& 24.5& 32.8& 24.3& 16.7& & 25.6
\\
		\hline
		Echo decay time $T_2$ at idle frequency ($\mu$s)& 8.2& 5.0& 6.1& 7.6& 8.1& 6.2& 2.6& 6.6& & 6.3
\\
		\hline
		Ramsey decay time $T_2^*$ at idle frequency ($\mu$s)& 4.5& 4.3& 5.2& 5.6& 5.3& 3.9& 2.5& 4.9& & 4.5
\\
		\hline
		Readout	$F_{00}$ (\%)& 97.7& 97.8& 96.9& 98.1& 96.7& 97.0& 98.5& 98.5& & 97.6
\\
		\hline
		Readout $F_{11}$ (\%)& 92.8& 91.8& 91.6& 91.3& 89.2& 93.5& 88.6& 92.5& & 91.4

		\\
		\hline
		Readout linewidth, $\kappa$/2$\pi$ (MHz)& 1.01& 1.08& 1.53& 0.55& 0.61& 1.87& 0.85& 0.76& & 1.03
\\
		\hline
		1Q XEB $e_{1}$ (\%))& 0.086& 0.081& 0.118& 0.122& 0.111& 0.087& 0.108& 0.082& & 0.099
\\
		\hline
		1Q XEB $e_{1}$ purity (\%)& 0.071& 0.083& 0.098& 0.104& 0.085& 0.080& 0.086& 0.077& & 0.085
\\
		\hline
		\hline
		Parameters& D1& D2& D3& D4& D5& D6& D7& D8& D9& Average\\
		\hline
		\hline
		Qubit maximum frequency, $\omega_{\textrm{q}}^{\textrm{max}}/2\pi$
		(GHz) & 5.024& 4.836& 4.852& 5.303& 5.180& 5.111& 5.245& 5.225& 5.206& 5.109
\\
		\hline
		Qubit idle frequency, $\omega_{\textrm{q}}/2\pi$ (GHz)& 4.858& 4.788& 4.746& 4.791& 4.806& 4.631& 4.829& 4.781& 4.843& 4.786
 \\
		\hline
		Readout drive frequency, $\omega_{\textrm{r}}/2\pi$ (GHz)& 6.382& 6.438& 6.491& 6.380& 6.435& 6.489& 6.373& 6.428& 6.482& 6.433
 \\
		\hline
		Qubit anharmonicity, $U/2\pi $ (MHz)& -250& -248& -252& -248& -248& -254& -247& -248& -254& -250
  \\
		\hline
		Energy relaxation time $T_1$ at idle frequency ($\mu$s)& 15.6& 31.9& 28.8& 31.8& 21.2& 40.8& 28.6& 34.8& 22.8& 28.5
\\
		\hline
		Echo decay time $T_2$ at idle frequency ($\mu$s)& 6.2& 8.3& 7.6& 3.5& 4.1& 3.8& 4.5& 2.5& 5.1& 5.1
\\
		\hline
		Ramsey decay time $T_2^*$ at idle frequency ($\mu$s)& 3.5& 6.3& 5.3& 1.5& 2.0& 1.5& 1.8& 1.2& 2.1& 2.8
\\
		\hline
		Readout	$F_{00}$ (\%)& 97.4& 96.6& 95.6& 97.5& 98.2& 97.4& 97.6& 97.7& 97.8& 97.3
 \\
		\hline
		Readout $F_{11}$ (\%)& 95.8& 94.8& 95.5& 96.3& 96.6& 93.7& 95.6& 95.5& 96.0& 95.5
\\
		\hline
		Readout linewidth, $\kappa$/2$\pi$ (MHz)& 0.99& 1.16& 0.33& 0.96& 0.47& 0.36& 1.22& 0.77& 0.59& 0.76
 \\
		\hline
		1Q XEB $e_{1}$ (\%)& 0.089& 0.102& 0.082& 0.099& 0.090& 0.095& 0.075& 0.080& 0.122& 0.093
\\
		\hline
		1Q XEB $e_{1}$ purity (\%)& 0.082& 0.113& 0.077& 0.095& 0.080& 0.091& 0.067& 0.067& 0.102& 0.086
\\
		\hline
		\hline
	\end{tabular}
	\caption{\textbf{Summary of single-qubit parameters.}}
	\label{tableSummary}
\end{table*}

\begin{table*}[htb!]
	\centering
	\begin{tabular}{c| c  c c c c c c}
		\hline
		\hline
		Apattern& D2-Z1& D3-X2& D4-Z3& D5-X3& D6-Z4& D9-X4& Average\\
		\hline
		2Q XEB $e_{2}$ (\%)& 1.78& 0.45& 0.82& 1.35& 0.52& 0.45& 0.90
\\
		\hline
		\hline
		Bpattern& D2-X2& Z1-D5& D4-X3& Z3-D7& Z4-D9& D8-X4& Average\\
		\hline
		2Q XEB $e_{2}$ (\%)& 0.68& 1.15& 1.63& 0.98& 1.08& 0.72& 1.04
\\
		\hline
		\hline
		Cpattern& X1-D2& D1-Z1& D3-Z2& X2-D6& D5-Z4& X3-D8& Average\\
		\hline
		2Q XEB $e_{2}$ (\%)& 1.28& 1.51& 0.74& 0.82& 1.49& 1.35& 1.20
\\
		\hline
		\hline
		Dpattern& X1-D1& Z1-D4& X2-D5& Z2-D6& X3-D7& Z4-D8& Average\\
		\hline
		2Q XEB $e_{2}$ (\%)& 0.86& 1.22& 0.74& 0.61& 1.12& 0.87& 0.90
\\
		\hline
		\hline
	\end{tabular}
	\caption{\textbf{Summary of CZ gates parameters.}}
	\label{tabletwoqubitgatesSummary}
\end{table*}

\begin{figure}[!htbp]
\begin{center}
\includegraphics[width=1.0\linewidth]{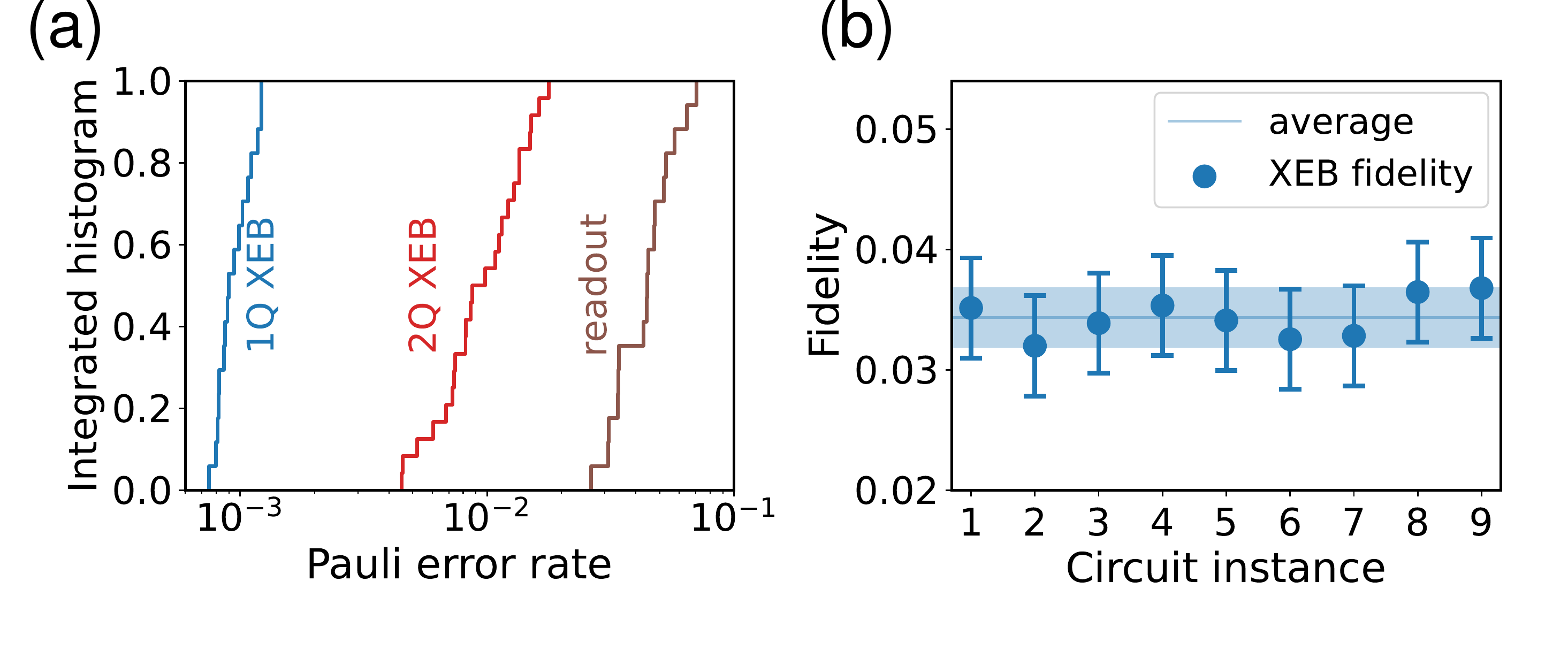}
\end{center}
\setlength{\abovecaptionskip}{0pt}
\caption{\textbf{System calibration}. \textbf{(a)} Integrated histogram of fidelities of single qubit $R_Y(\pi/2)$ rotation (1Q XEB), CZ gate (2Q XEB) and readout error (readout). 
\textbf{(b)} Cross entropy benchmarking fidelity for 9 instances of 17-qubit 20-cycle random circuits. Error bars describes $\pm 5\sigma$ statistical deviation, color band is for $\pm 3\sigma$ with $\sigma= 1/\sqrt{N_{\text{sample}}}$, where $N_{\text{sample}}=1,440,000$.}
\label{sfig1}
\end{figure}

The basic calibration steps for the employed 17 qubits, 24 couplers, 7 readout resonators and 7 JPAs are the same with
\textit{Zuchongzhi} 2.0. One can see more details about the procedure in Ref~\cite{wu2021strong}. The fine calibration steps are the same with Ref~\cite{zhao2022}. 
The device parameters are summarized in Table \ref{tableSummary} and Table
\ref{tabletwoqubitgatesSummary}. Fig.~\ref{sfig1}(a) displays the integrated histogram for single qubit gate (25ns duration, average error of $0.096\%$), CZ gate (32ns duration, average error of $1.009\%$) and readout (1.5$\mu s$ duration + 2.4$\mu s$ idle operation, average error of $4.469\%$) error after calibration.

We also perform 17-qubit 20-cycle random circuit sampling to benchmark the performance of whole system (see Fig.~\ref{sfig1}(b)), similar to the previous work~\cite{zhao2022}. The average fidelity achieved on 9 different instances of random circuits  is $0.0341\pm0.0018$, which is in high agreement with the predicted value 0.0343 obtained by taking the product of the Pauli fidelity of each single-qubit, two-qubit and measurement operation.

\section{Thermal excitation mitigation and Readout optimization}

\begin{figure}[!htbp]
\begin{center}
\includegraphics[width=1.0\linewidth]{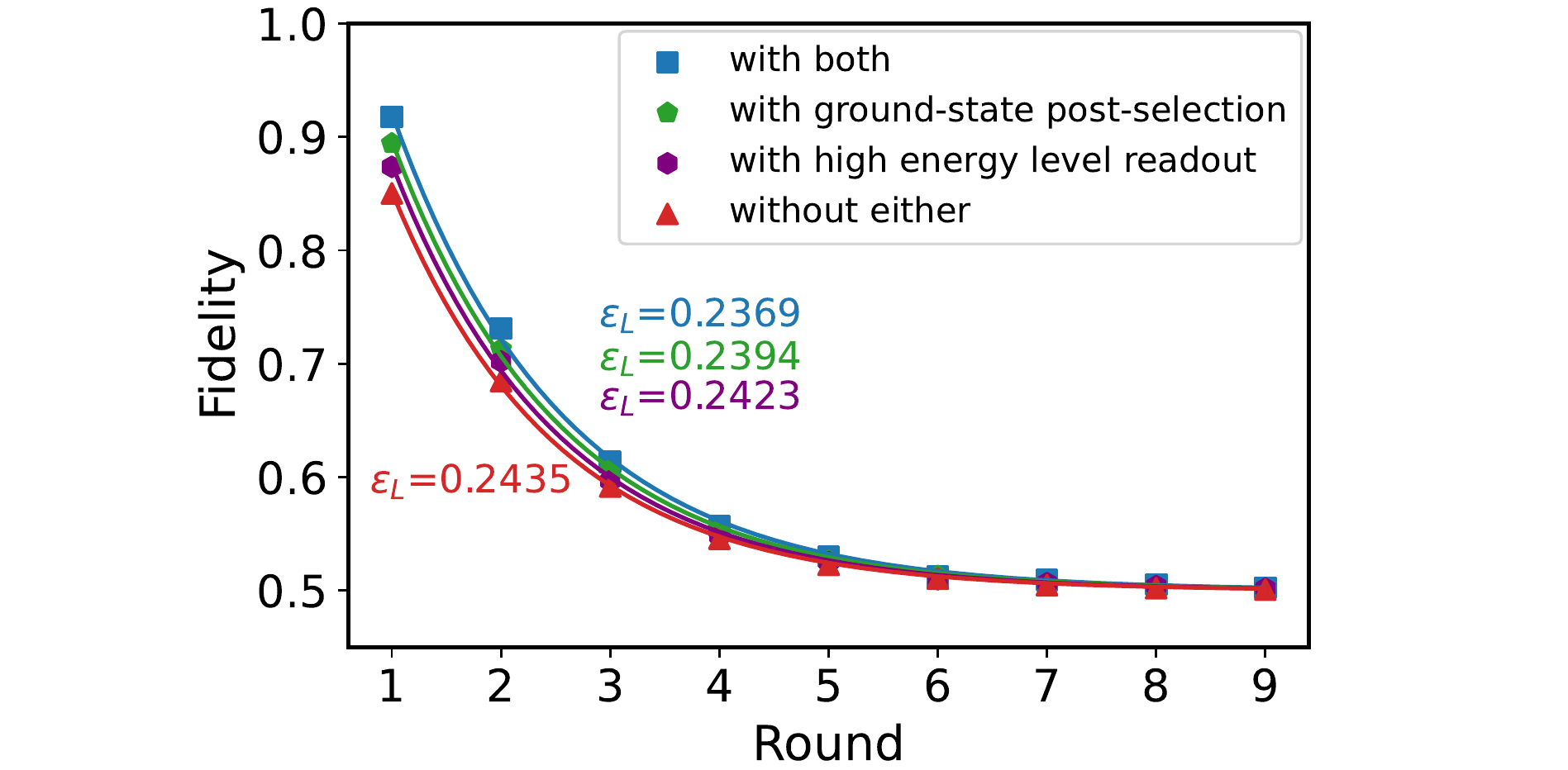}
\end{center}
\setlength{\abovecaptionskip}{0pt}
\caption{\textbf{Comparison of using ground-state post-selection and high energy level readout}. For the $|0\rangle_L$ state, multiple rounds of error correction cycles are applied. The error corrected logical fidelities and logical error rates per round when using no, using one of, and using both ground-state post-selection and high energy level readout scheme are shown.}
\label{sfig2}
\end{figure}

To reduce thermal excitation errors, the ground-state post-selection scheme~\cite{krinner2022realizing} is applied, \textit{i.e.} one round of readout is performed for all qubits before running the quantum circuit, and drop the results where thermal excitation occurred. In our experiments, a retention rate of $57.17\pm0.02\%$ is obtained, which corresponds to an average thermal excitation rate of $3.235\pm0.002\%$ per qubit.

To improve the readout fidelity, the high energy level readout scheme is applied to the data qubits, \textit{i.e.} driving the qubit to higher levels during the readout~\cite{Elder_2020}.

In Fig.~\ref{sfig2}, we experimentally investigate the effects of ground-state post-selection and high energy level readout to the error correction performance. It can be seen that both schemes can improve the fidelity of the logical state and slightly reduce the logical error rate per round. And the combination of these two schemes can get better results, so we used them in our experiments.


\section{Physical implementation of arbitrary single-qubit state preparation}

An arbitrary single-qubit pure state $|\psi\rangle$ can be represented as a point on the surface of Bloch sphere:
\begin{align}
|\psi\rangle = \cos{\left(\frac{\theta}{2}\right)}|0\rangle + e^{i\varphi} \sin{\left(\frac{\theta}{2}\right)} |1\rangle
\end{align}
where $\theta$ and $\varphi$ denote the polar and azimuthal angles of the point on Bloch sphere, respectively. The pure state $|\psi\rangle$ can be realized from $|0\rangle$ by two rotations around the coordinate axis,
\begin{align}
|\psi\rangle = R_z(\varphi+\pi/2) \cdot R_x(\theta) |0\rangle
\end{align}

To improve the fidelity of the arbitrary rotation gate and simplify the calibration procedure, we have followed the scheme in Ref.~\cite{McKay2017}, the arbitrary rotation around the $X$-axis is represented as a combination of $Z$ and $X_{\pi/2}$ gates,
\begin{align}
R_x(\theta) = Z_{-\pi/2} \cdot X_{\pi/2} \cdot Z_{\pi-\theta} \cdot X_{\pi/2} \cdot Z_{-\pi/2},
\end{align}
thus, the pure state $|\psi\rangle$ can be prepared as
\begin{align}
|\psi\rangle = Z_{\varphi} \cdot X_{\pi/2} \cdot Z_{\pi-\theta} \cdot X_{\pi/2}|0\rangle.
\label{prepare}
\end{align}
Here the phase gate acting on the $|0\rangle$ state is ignored, and the $Z$ gate is implemented as the virtual-$Z$ gate in experiments, which is basically perfect. Thus, we only need to calibrate one $X_{\pi/2}$ gate to prepare an arbitrary high-fidelity quantum state.

\section{Logical \textit{Y} measurement in the error correction procedure}

There are two approaches for applying logical $Y$ measurements to surface codes: (1) Measuring logical $X$ operator followed by a logical $S$ gate; (2) Measuring the logical $Y$ operators directly. The first of these approaches is fault-tolerant, while the second is not fault-tolerant. However, the implementation of the logical $S$ gate requires the use of magic state injection protocol, which is currently impractical.

For the second approach, the measurement of logical operator $\hat{Y}_L = +i \hat{X}_L \hat{Z}_L$, is the computed product of data-qubit outcomes 
\begin{align} \label{eq:a1}
Y_L = Y_{D_c}\prod_{D_i\in \{X_L\}\setminus \{Z_L\}} X_{D_i} \prod_{D_j\in \{Z_L\}\setminus \{X_L\}} Z_{D_j},
\end{align}
where $D_c = \{X_L\} \cap \{Z_L\}$, and the sets $\{X_L\}$ and $\{Z_L\}$ represent the data qubits belonging to the logical operators $\hat{X}_L$ and $\hat{Z}_L$, respectively. That is, the logical $Y$ measurement is equivalent to measure the data qubit $D_c$ in the $Y$ basis, and the other data qubits belonging to $\hat{X}_L$ in the $X$ basis and $\hat{Z}_L$ in the $Z$ basis.


For the measurement of logical $X$ ($Z$) operator, all data qubits are measured in $X$ ($Z$) basis, thus allowing the calculation of the values of all $X$- ($Z$-) stabilizers, which are key to fault-tolerant. However, this method can not work for the measurement of the logical $Y$ operator. Here, we proposed a measurement approach for data qubits in logical $Y$ operator measurement, which is able to calculate half of the $X$-stabilizers and half of the $Z$-stabilizers for error correction, and has been used in the error correction experiments in the main text.

\begin{figure}[!htbp]
\begin{center}
\includegraphics[width=1.0\linewidth]{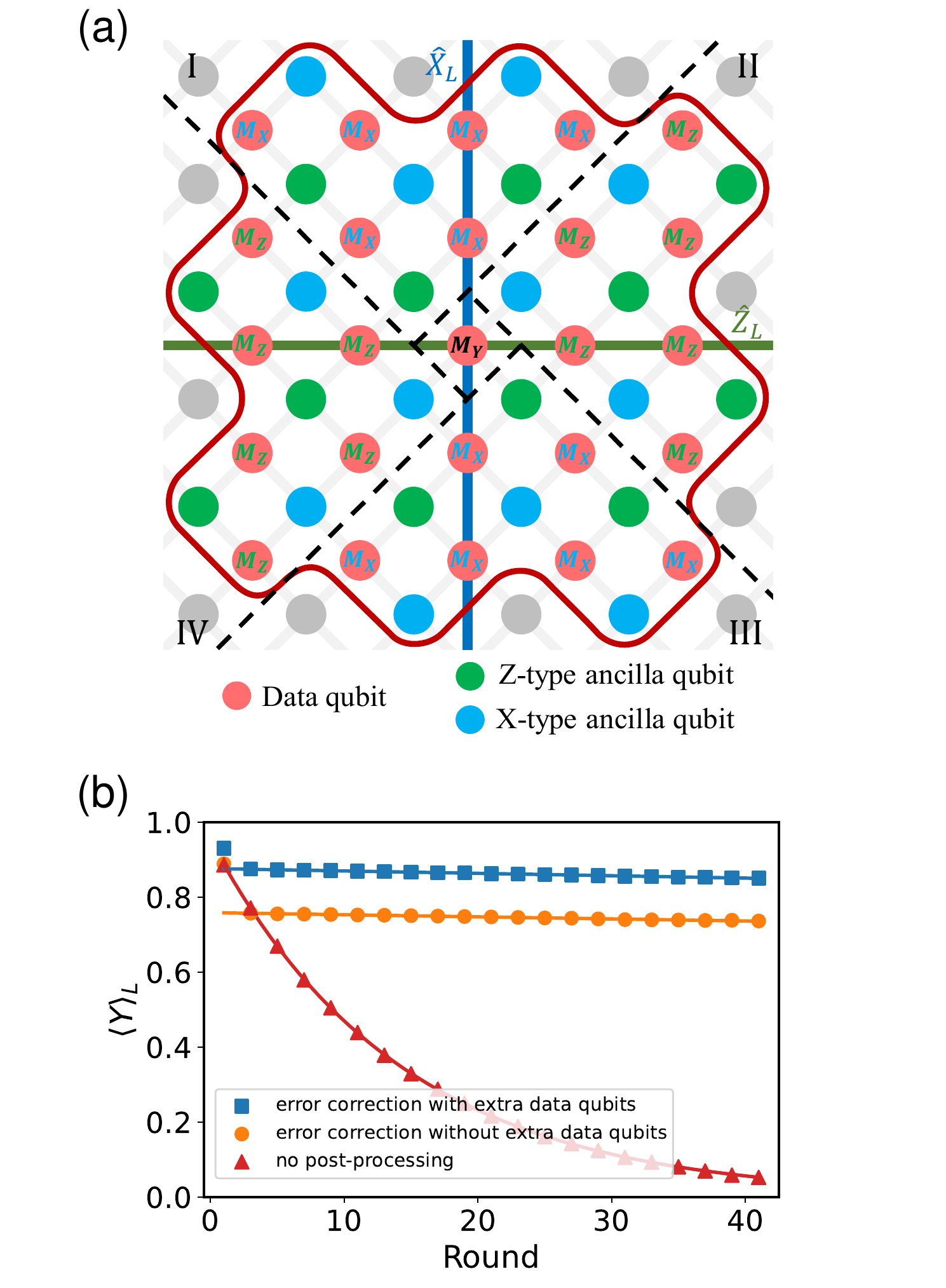}
\end{center}
\setlength{\abovecaptionskip}{0pt}
\caption{\textbf{Logical $Y$ measurement and numerical simulation results}. \textbf{(a)} The logical $Y$ measurement in a distance-5 surface code. Data qubits in regions I, III (II, IV) are measured in the $X$ ($Z$) basis, and data qubit in the center is measured in the $Y$ basis. \textbf{(b)} Simulation results of the logical $Y$ measurement. The logical state ${|{+i}\rangle}_L$ is prepared in a distance-5 surface code and then measured in logical $Y$ basis with different error correction cycles. The orange and blue dots indicate the logical excitation with error correction by using or not using the extra data qubits, and all lines are fitted with exponential decay function.} 
\label{sfig3}
\end{figure}

The measurement of data qubits is shown in Fig.~\ref{sfig3}(a),
\textit{i.e.} in addition to measuring the data qubits mentioned above, all the other data qubits in regions I and III (II and IV) are measured in the $X$ ($Z$) basis. In this way, the $X$-stabilizers in regions I, III and $Z$-stabilizers in regions II, IV are determined by the measurement results of the data qubits.


We study the performance of the logical $Y$ measuruement by numerical simulation, the logical expectation results are shown in Fig.~\ref{sfig3}(b). The average physical error rate is set to $0.1\%$, logical state $|{+i}\rangle_L$ is prepared in a distance-5 rotated surface code and be measured in the logical $Y$ basis with different error correction cycles. We compare the error correction performance of logical $Y$ measurement with and without the extra data qubits.
As you can see, the extra data qubits can effectively reduce the error rate of non-fault-tolerant logical $Y$ measurement.

\section{Experimental results of preparing different logical states}

\begin{figure*}[!htbp]
\begin{center}
\includegraphics[width=1.0\linewidth]{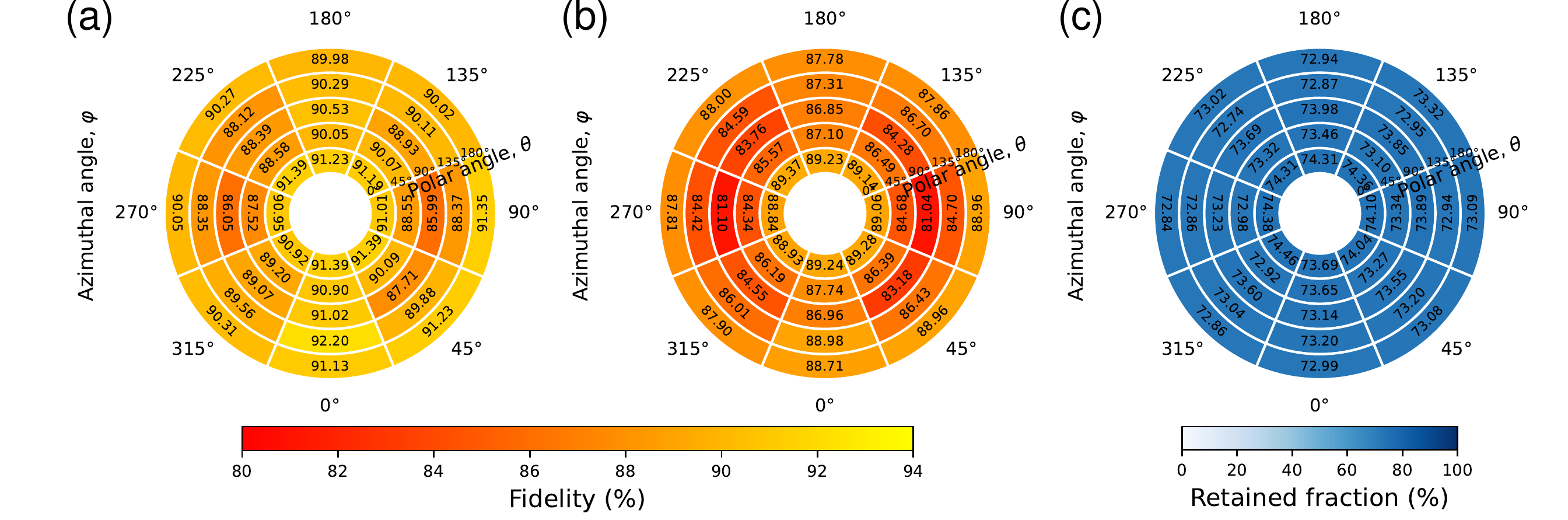}
\end{center}
\setlength{\abovecaptionskip}{0pt}
\caption{\textbf{Logical fidelities of the prepared logical states in experiment}. \textbf{(a)} and \textbf{(b)}  Logical fidelity of preparation at different locations on the bloch sphere with and without post-selection, respectively. We represent the bloch sphere as a pie shape, and each annular sector on the pie refers to a point on the bloch sphere, with the radial direction representing the polar angle $\theta$ and the tangential direction representing the azimuthal angle $\varphi$ (we use the angle system here for convenience). \textbf{(c)} The average retained fraction of post-selection for different logical states.}
\label{sfig6}
\end{figure*}

In Fig.~\ref{sfig6}(a) and (b), we show the logical fidelities of preparation different logical states with or without post-selection. The average fidelities in these two caes are \hhl{$0.8983\pm0.0002$ and $0.8671\pm0.0002$}, respectively. Fig.~\ref{sfig6}(c) is the retained fraction of the post-selection, with an acceptable mean value of $73.413\pm0.008\%$. It can be seen that for different logical states, the post-selection retained fraction is almost the same.

\section{Detection event fraction and error detection correlation}
\vspace{-0.3cm}

\begin{figure}[!htbp]
\begin{center}
\includegraphics[width=1.0\linewidth]{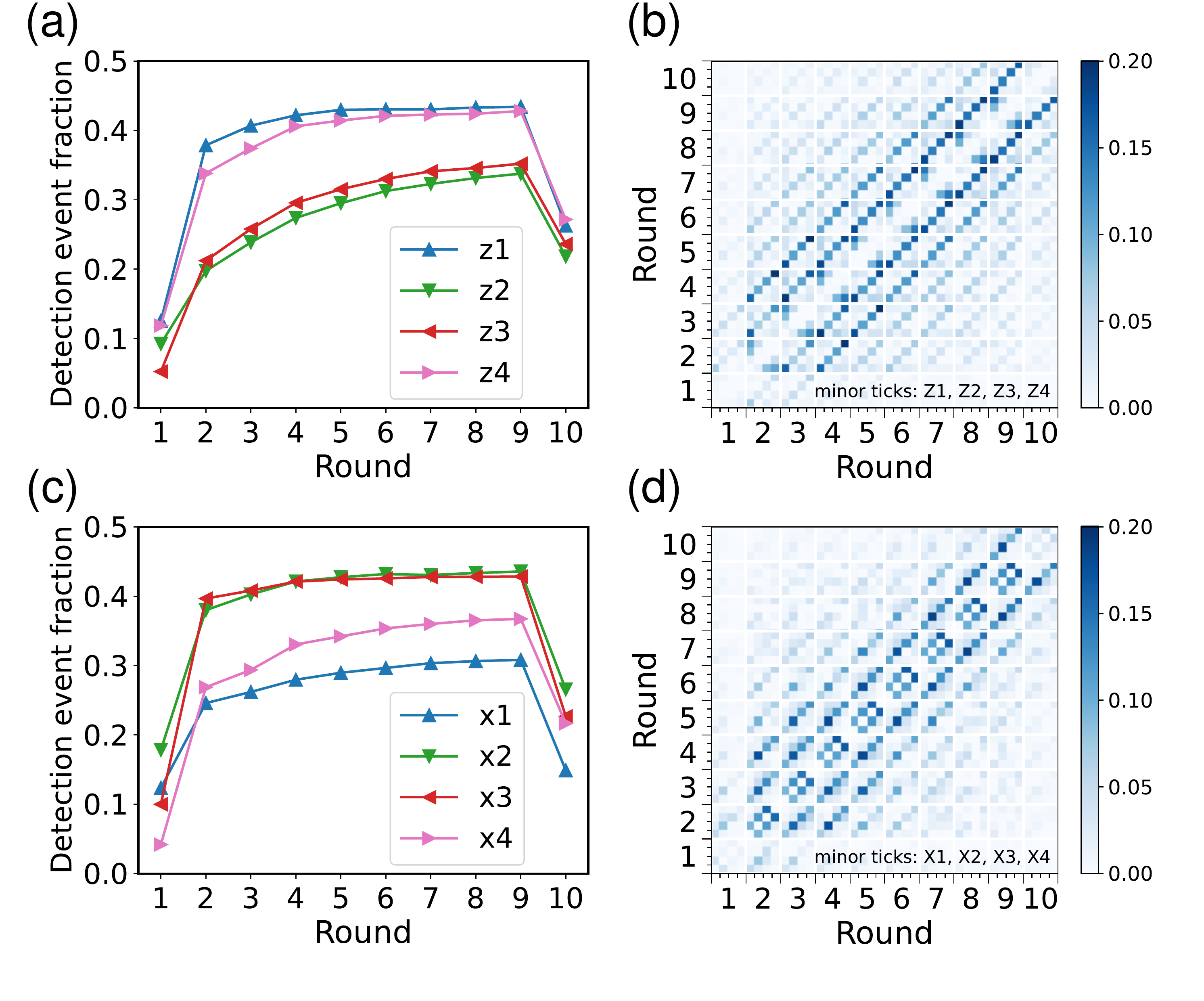}
\end{center}
\setlength{\abovecaptionskip}{0pt}
\caption{\textbf{Detection event fraction and correlation matrix}. 
\textbf{(a)} Detection event fraction (DEF) for the $|{0}\rangle_L$ state. Each line corresponds to the DEF for one $Z$-stabilizer, and the last round is obtained by comparing the results of the data qubits measurements with the last two rounds of syndrome measurements. \textbf{(b)} Correlation matrix in space-first node ordering for the $|{0}\rangle_L$ state. The major ticks indicate different surface code cycles, and the minor ticks indicate various stabilizers, we have truncated the correlation matrix at $p=0$ to ignoring higher order correlations. \textbf{(c), (d)} Detection event fraction and Correlation matrix for the $|{-}\rangle_L$ state.}
\label{sfig7}
\end{figure}

We prepared the $|{0}\rangle_L$ ($|{-}\rangle_L$) state using the standard state preparation scheme and performed up to 9 rounds of surface code cycles, and analyzed the errors occurred in the cycles of the surface code using the detection event fraction and correlation matrix methods~\cite{ai2021exponential}.

We shows the detection event fraction of $Z$-stabilizers in Fig.~\ref{sfig7}(a). Here, the last round is obtained by comparing the results of the data qubits measurements with the last two rounds of syndrome measurements,  the values of it is apparently lower than the others is a result of skipping idle operation which applies to all measurements. Since the first round data is copied directly from the measurement without comparing with a physical data, the influence of an idle operation is also reduced. Besides the first and last rounds, the fraction of detection events increases with the number of surface code rounds, showing the accumulation of leakage. The values of $Z_1$, $Z_4$ stabilizers are significantly higher than the other two stabilizers because both of them are wight-4 stabilizers (the other are wight-2). The detection event fraction of $X$-stabilizers is shown in Fig.~\ref{sfig7}(c), which the result is mostly consistent with the $Z$-stabilizers.

The Fig.~\ref{sfig7}(b) shows the correlation matrix for the $Z$-stabilizers. The major and minor ticks in this figure indicate different surface code cycles and various stabilizers, respectively. We have truncated the correlation matrix at $p=0$ to ignoring higher order correlations. Each block represents the correlation between two points, and a larger value means a stronger correlation between these two. As can be seen from the figure, the S-edges are almost invisible, but the well-visible T-edges indicate the presence of long-time correlations in detection events lasting for over 6 rounds. 
For the $X$-stabilizers which is shown in Fig.~\ref{sfig7}(d), there is the same long-time correlations lasting for over 5 rounds, However, the S-edges are well-visible.


\section{Analysis of logical error for the state preparation scheme}

\begin{figure}[!htbp]
\begin{center}
\includegraphics[width=1\linewidth]{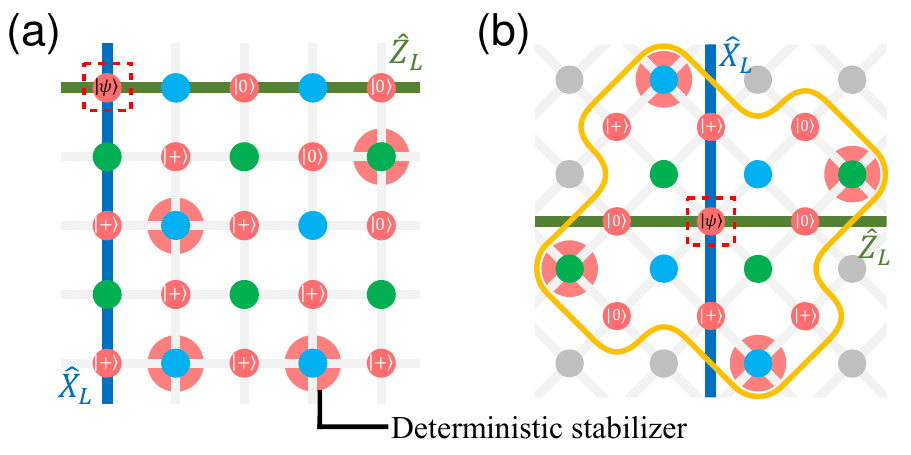}
\end{center}
\setlength{\abovecaptionskip}{0pt}
\caption{\textbf{Comparison of different schemes.} \textbf{(a)} Li's scheme for the regular distance-three surface code. \textbf{(b)} our scheme for the rotated distance-three surface code. The magic-state qubit in the center. The markers on the data qubits indicate the states to be prepared, and the pink background around the ancilla qubits represent the deterministic stabilizers in the state preparation.}
\label{sfig5}
\end{figure}

In Ref.~\cite{li2015magic}, Li has proposed a scheme to prepare logical magic state on the regular surface code. Take the distance-three as an example, the data qubit initialization strategy is shown in Fig.~\ref{sfig5}(a). The markers on the data qubits indicate the states to be prepared, and the data qubit in the top-left corner (called magic-state qubit thereafter) is prepared to the target state. This scheme has 4 deterministic stabilizers (indicated by the pink background in the Fig.~\ref{sfig5}(a)) that can be used to detect errors in the first surface code cycle. The logical error rate during state preparation can be suppressed using the post-selection method. However, some undetectable errors can lead to logical errors.

In contrast, our scheme (Fig.~\ref{sfig5}(b)) is used for the rotated surface code and places the magic-state qubit at the center of the surface code, which has the advantage of having more deterministic stabilizers: half of the $X$-stabilizers and half of the $Z$-stabilizers for an arbitrary surface code distance (4 deterministic stabilizers in distance-3 surface code).

To estimate logical error rate, we follow the method in the Li's paper, consider only circuit-level noise, and use the depolarizing error model, which means: 1.Attempting to perform single-qubit gate, but performing in addition one of the single-qubit operations $X$, $Y$, $Z$ with probability $p_1/3$ each. 2. Attempting to perform CNOT (CZ) gate, but performing in addition one of the two-qubit operations ${\{I, X, Y, Z\}}^{\otimes 2}\setminus\{I\otimes I\}$ with probability $p_2/15$ each. 3. Attempting to initialize qubit to $|0\rangle$ but instead to $|1\rangle$ with probability $p_I$. 4. Attempting to measure a qubit in $Z$ basis, but getting the wrong value and projecting to the wrong state with probability $p_M$.

In the case of lower physical gate error rates, only the first-order errors of $p$ are considered. The logical errors here are those that cause logical measurement errors (including $\hat{X}_L$, $\hat{Y}_L$, $\hat{Z}_L$) but do not generate detection events. We analyzed and calculated the logical error rate in different cases with the support of the stim tool~\cite{gidney2021stim}, and denoted the logical error rate as
\begin{align}
p_L = a \cdot p_1 + b \cdot p_2 + c \cdot p_I + \mathcal{O}(p^2).
\end{align}
where the parameters are listed in Table~\ref{cnot error} (use CNOT gate) and Table~\ref{cz error} (use CZ gate).

In the case of using CNOT gates, the single-qubit gate errors that would cause a logical error occur only during the preparation of the magic-state qubit to the magic state and and do not change with the code distance grows. For the $H$-type magic state, the parameter $a = 2/3$, while for the $T$-type magic state, since an additional single-qubit gate is required (see Eq.~\ref{prepare}), the parameter $a$ would be $5/3$. However, using CZ gates, errors at single-qubit gate which used to form the surface code cycle may also lead to logical errors since the stabilizers that can be used to detect errors during the state preparation is not complete.

We also compare the logical error rate of post-selection with only one and two round(s) of surface code cycle(s), and it is clear that two rounds have an advantage over only one round. However, when the physical gate error rate is not low enough, the success rate of the post-selection using two rounds of surface codes will be much lower than using only one round, and the magic state distillation protocol requires a large number of raw logical magic states, so in some cases using only one round will be more advantageous.

\begin{table}[!htb]
	\renewcommand{\arraystretch}{1.8}
	\centering
	\begin{tabular}{p{0.7cm}|p{1cm}|p{2.4cm}|p{1cm}|p{2.4cm}}
		\hline
		\hline
		& \multicolumn{2}{c|}{round $=1$} & \multicolumn{2}{c}{round $=2$} 
	\\ 
		\cline{2-5}
	    & \makecell[c]{$d=3$} & \makecell[c]{$d>3$ ($d$ is odd)} &  \makecell[c]{$d=3$} & \makecell[c]{$d>3$ ($d$ is odd)}
	\\ 
		\hline 
		\hline
		\makecell[c]{$a$} & \multicolumn{4}{c}{\makecell[c]{
			$2/3$ ($H$-type) \\
			$5/3$ ($T$-type)
			}
		}
	\\
		\hline 
		\makecell[c]{$b$} & \makecell[c]{$94/15$} & \makecell[c]{$38/15+4d/3$} & \multicolumn{2}{c}{$2/15+4d/5$}
	\\ 
		\hline
		\makecell[c]{$c$} & \multicolumn{4}{c}{$1$}
	\\
		\hline
		\hline
	\end{tabular}
	\caption{\textbf{Logical error rate with CNOT gate.}}
	\label{cnot error}
\end{table}

\begin{table}[!htb]
	\renewcommand{\arraystretch}{1.8}
	\centering
	\begin{tabular}{p{0.7cm}|p{1cm}|p{2.4cm}|p{1cm}|p{2.4cm}}
		\hline
		\hline
		& \multicolumn{2}{c|}{round $=1$} & \multicolumn{2}{c}{round $=2$} 
	\\ 
		\cline{2-5}
	    & \makecell[c]{$d=3$} & \makecell[c]{$d>3$ ($d$ is odd)} &  \makecell[c]{$d=3$} & \makecell[c]{$d>3$ ($d$ is odd)}
	\\ 
		\hline 
		\hline
		\makecell[c]{$a$} & \multicolumn{2}{c|}{\makecell[c]{
				$10/3+4d/3$ ($H$-type) \\
				$13/3+4d/3$ ($T$-type)
				}
			} & \multicolumn{2}{c}{\makecell[c]{
				$4/3+d$ ($H$-type) \\
				$7/3+d$ ($T$-type)
				}
			}
	\\
		\hline 
		\makecell[c]{$b$} & \makecell[c]{$94/15$} & \makecell[c]{$38/15+4d/3$} & \multicolumn{2}{c}{$2/15+4d/5$}
	\\ 
		\hline
		\makecell[c]{$c$} & \multicolumn{4}{c}{$1$}
	\\
		\hline
		\hline
	\end{tabular}
	\caption{\textbf{Logical error rate with CZ gate.}}
	\label{cz error}
\end{table}

\bibliography{references}